\documentclass{emulateapj}

\def\msun{$M_{\odot}$ }

\def\mult{$\times$}

\begin{document}
  \title{Active Carbon and Oxygen Shell Burning Hydrodynamics}
  \author{Casey A. Meakin\altaffilmark{1} \& David Arnett\altaffilmark{1} }
  \altaffiltext{1}{Steward Observatory, University of Arizona, Tucson, AZ 85721}
  \email{cmeakin@as.arizona.edu, darnett@as.arizona.edu}
  
\begin{abstract}
  We have simulated 2.5$\times$10$^3$ seconds of the late evolution of a $23 \rm M_\odot$ 
  star with full hydrodynamic behavior.   
  We present the first simulations of a multiple-shell burning epoch, including the 
  concurrent evolution and interaction of an oxygen and carbon burning shell. 
  In addition, we have evolved a 3D model of the oxygen burning shell to sufficiently 
  long times (300 seconds) to begin to assess the adequacy of the 2D approximation.
  We summarize striking new results: (1) strong interactions occur between active carbon and 
  oxygen burning shells, (2) hydrodynamic wave motions in nonconvective regions,
  generated at the convective-radiative boundaries, are energetically important in
  both 2D and 3D with important consequences for compositional mixing, and 
  (3) a spectrum of mixed p- and g-modes are unambiguously identified with corresponding 
  adiabatic waves in these computational domains.  
  We find that 2D convective motions are exaggerated relative to 3D because of vortex 
  instability in 3D. We discuss the implications for supernova progenitor evolution and 
  symmetry breaking in core collapse.
\end{abstract}

\keywords{hydrodynamics, turbulence, 
  stars: interiors, core collapse}

\section{Introduction}

\par Numerical simulations of stellar evolution are generally based
upon restrictive assumptions regarding dynamics (e.g., hydrostatic
balance and mixing-length convection), because the dynamic timescales are so
much shorter than the nuclear burning timescale. Neutrino cooling
accelerates the last burning stages so that direct dynamic simulation is feasible
\citep{bazan1998,asida2000}, at least for the oxygen burning shell. 
We are extending this work to longer evolutionary times,
larger computational domains, and three dimensional flow (3D). 
In this letter, we summarize new results with a discussion of the hydrodynamics
underlying important symmetry breaking and compositional mixing processes which
may significantly affect progenitor and core-collapse supernova models.
A detailed discussion of these results will appear separately;
we indicate the wider implications here.

\section{A Double Shell Model: Active Oxygen and Carbon Burning}

\par
\begin{figure}
  \plotone{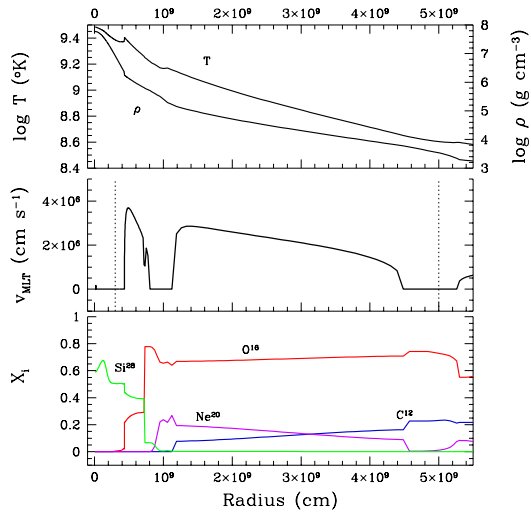}
  \caption{\label{initmodel}
    The 23\msun TYCHO model used as initial conditions for the hydrodynamic
    simulation is shown here.  ({\em top}) Density and temperature profiles 
    are shown illustrating the large number of scale heights simulated
    as well as the complex entropy structure due to the burning shells.
    ({\em middle})  The mixing length velocities are shown and delineate the 
    extents of the carbon and oxygen burning convection zones. The 
    dashed vertical lines mark the boundaries of the simulation domain.
    ({\em bottom}) The onion skin compositional layering of the model is shown here
    for the four most abundant species.
    \label{f1}}
\end{figure}

 Previously we have evolved a 23\msun model with the one-dimensional
TYCHO code to  a point where oxygen and carbon are burning in 
concentric convective shells which overlay a silicon-rich core;
details may be found in \citep{young2005a}.
The resultant stellar structure is presented in Figure \ref{f1}.
For subsequent hydrodynamic evolution we use PROMPI, a 
version of the PROMETHEUS direct Eulerian PPM code \citep{fryxell1989} that has 
been ported to multi-processor computing systems via domain decomposition.  
Our multi-dimensional calculations use the same physics as the one-dimensional
TYCHO code, including nuclear reaction rates, equation of state, 
and radiative opacities.  We use a 25 nucleus reaction network in these
models, tuned to capture the oxygen and carbon burning energy
generation rates to within 1\% of the 177 species version used to
evolve the 1D model.
The 25 nucleus network contains electrons, neutrons, protons, $^4\rm He$, $^{12}\rm C$,
$^{16}\rm O$, $^{20}\rm Ne$, $^{23}\rm Na$, 
$^{24}\rm Mg$, $^{28}\rm Si$, $^{31}\rm P$, 
$^{32}\rm S$, $^{34}\rm S$,  $^{35}\rm Cl$,  $^{36}\rm Ar$,  
$^{38}\rm Ar$, $^{39}\rm K$,  $^{40}\rm Ca$,  $^{42}\rm Ca$,  
 $^{44}\rm Ti$,  $^{46}\rm Ti$,  $^{48}\rm Cr$,  $^{50}\rm Cr$,
  $^{52}\rm Fe$, $^{54}\rm Fe$, $^{56}\rm Ni$,
and all significant strong and weak interaction links. The reaction
rates, including $\rm ^{12}C(\alpha,\gamma)$, are from \citet{rt00}.

\par  A two dimensional model has been calculated on a  $90^\circ $ wedge
which is embedded in the equatorial plane of a spherical coordinate system and has
radial limits which encompass both the oxygen and carbon burning convective shells.
Table 1 lists some additional details of the 
simulated model.  A three dimensional model including just the oxygen
shell and bounding stable layers is partially evolved at present (300 seconds).

\section{Results}

\par
\begin{figure}
  \plotone{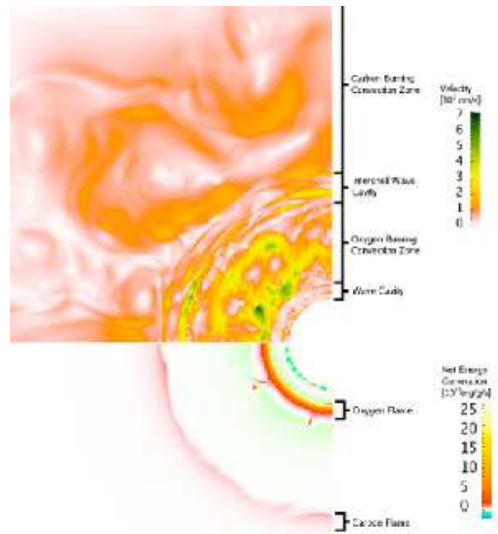}
  \caption{
    The magnitude of the flow velocity ({\em top}) and the net energy generation rate, 
    $\epsilon_{net} = \epsilon_{nuc} + \epsilon_{\nu}$  ({\em bottom}), is shown for 
    a snapshot of the simulation which includes both an oxygen and carbon burning 
    convection zone.
    The carbon burning convective shell extends to a radius of $\sim$4.5\mult10$^9$ cm 
    while the figure is truncated at a radius of $\sim$2.9\mult10$^9$ cm for clarity.  
    A weak silicon-burning convection zone develops at the inner edge of the grid due 
    to a small  boundary-zone entropy error which accumulates during the course 
    of the calculation.
    \label{f2}}
\end{figure}

 {\em Flow Topology with Two Burning Shells.}
Following the readjustment of the outer boundary due to small
inconsistencies in the initial 1D model, a quasi-steady
state flow develops, shown in Fig.\ref{f2}.
The top half of the figure shows the velocity magnitude; the lower
shows energy generation. {\em Velocities are significant even in the
nonconvective regions}, but have different morphology. The convective
regions have round patterns (vortices) with occasional plumes, 
while the nonconvective regions have flattened patterns (mostly g-modes). 
The flow fluctuates strongly. New fuel is ingested from above; the
oxygen flame shows ``feathery'' features corresponding to such
fuel-rich matter  flashing as it descends. This was previously seen
\citep{bazan1998,asida2000}. A new feature appears in the movie
version of Fig.~2, which shows a pronounced, low order distortion of
the comoving coordinate, squashing and expanding the apparent circles
on which carbon burning proceeds.
This is due to the coupling of the two shells by waves in the
nonconvective region between them. This behavior seems robust; we
expect it to persist, so that at core collapse this part of the star
(at least) will have significant nonspherical distortion.

\begin{deluxetable}{lll}  
 
\tablewidth{0pt}
\tablecolumns{3}

\tablecaption{Model Parameters}
\tablehead{
  \colhead{Quantity} & 
  \colhead{Value} \\}

\startdata
Stellar mass ($M_{\odot}$) \dotfill\dots           & 23\\
Stellar age (yr)      \dotfill\dots          & {2.3\mult10$^6$}\\
Oxygen shell convective timescale\tablenotemark{a} (s) \dotfill\dots & {$\sim$10$^2$}\\
Carbon shell convective timescale\tablenotemark{a} (s) \dotfill\dots & {$\sim$10$^3$}\\
Hydro simulation time (s)     \dotfill\dots        & 2.5\mult10$^3$   \\
Inner, outer grid radius (10$^9$ cm) \dotfill\dots & 0.3, 5.0         \\
Pressure scale heights across domain \dotfill\dots & $\sim$ 9         \\
Angular extent of grid (rad)         \dotfill\dots & $\pi$/2          \\
Grid zoning, n$_r$\mult n$_{\phi}$\mult n$_{\theta}$\dotfill\dots & 800\mult320\mult1 \\
Numer of timesteps                 \dotfill\dots  & $\sim$1.5\mult10$^6$  \\
\enddata

\tablenotetext{a}{These convective timescales are based on the mixing length theory
  velocities.}

\end{deluxetable}

\par
\begin{figure}
  \plotone{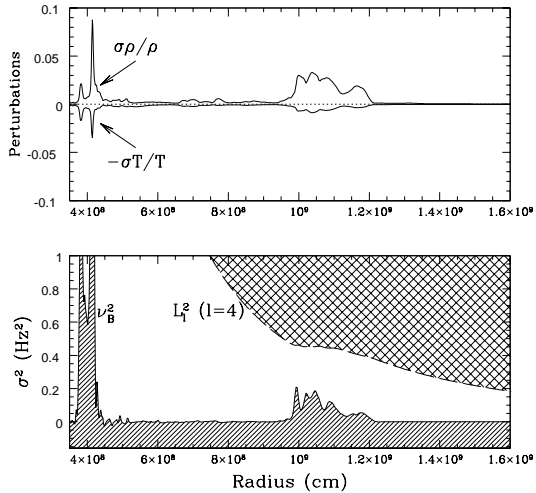}
  \caption{
    {\em (top)} Nonspherical density and temperature perturbations  
    (root mean square fluctuation about the angular mean divided by the angular mean) 
    are shown for the inner 1.2\mult10$^9$ cm of the simulation domain. 
    {\em (bottom)} In this propagation diagram (showing oscillation frequency versus 
    stellar radius) the gravity wave and (l=4) acoustic wave propagation zones are
    indicated by the horizontal and cross-hatched shading, respectively.
    The gravity wave cavity is bounded above by the buoyancy frequency (solid line), 
    while the acoustic cavity is bounded below by the Lamb frequency (dashed line).
    \label{f3}}
\end{figure}

{\em Stiffness and the Source of Density Perturbations.}
Another asymmetry, nonspherical density perturbations, was found by
\citet{bazan1998,asida2000}. 
The fluctuations in density and temperature,
presented in the top panel of Figure \ref{f3} as root mean square deviations 
from an angular mean,  reach values as large as $\sim$10\% 
and are localized at the nonconvective region just beyond the convective
boundary (top panel). 
The fluctuations are coincident with regions where the buoyancy frequency,
$\nu_B$, is large, which can be seen in the bottom panel of Figure \ref{f3}.
Here, $\nu_B^2$ is a measure of the ``stiffness'' of the
stratification \citep{turner73}, 
and is proportional to the restoring buoyancy force on perturbed stellar 
matter,

\begin{equation}
  \nu_B^2 = -g\Big(\frac{\partial\ln\rho}{\partial r}\Big|_s - \frac{d\ln\rho}{dr}\Big),
  \label{bv}
\end{equation}

\noindent where $g$ is the gravity, and the term in parentheses is the
difference between the fractional density gradient of the stellar structure and the 
fractional density change due to a radial (adiabatic) Lagrangian displacement.  Regions 
where $\nu_B$ is zero are unstable to convective motions. The spikes in $\nu_B$ in our 
model are due to steep, stabilizing composition gradients which separate fuel from ash 
and lead to sharp gradients in density.
Convection excites wave motions in the adjacent stable layers which give rise to the
density perturbations.  Similar internal wave phenomena can be observed in laboratory
ice-water convection experiments where the largest temperature fluctuations are 
measured immediately above the convecting layer where the buoyancy frequency is 
large \citep{townsend1966a}, highlighting the generality of this phenomenon.

\begin{figure*}
  \epsscale{1.1}
  \plottwo{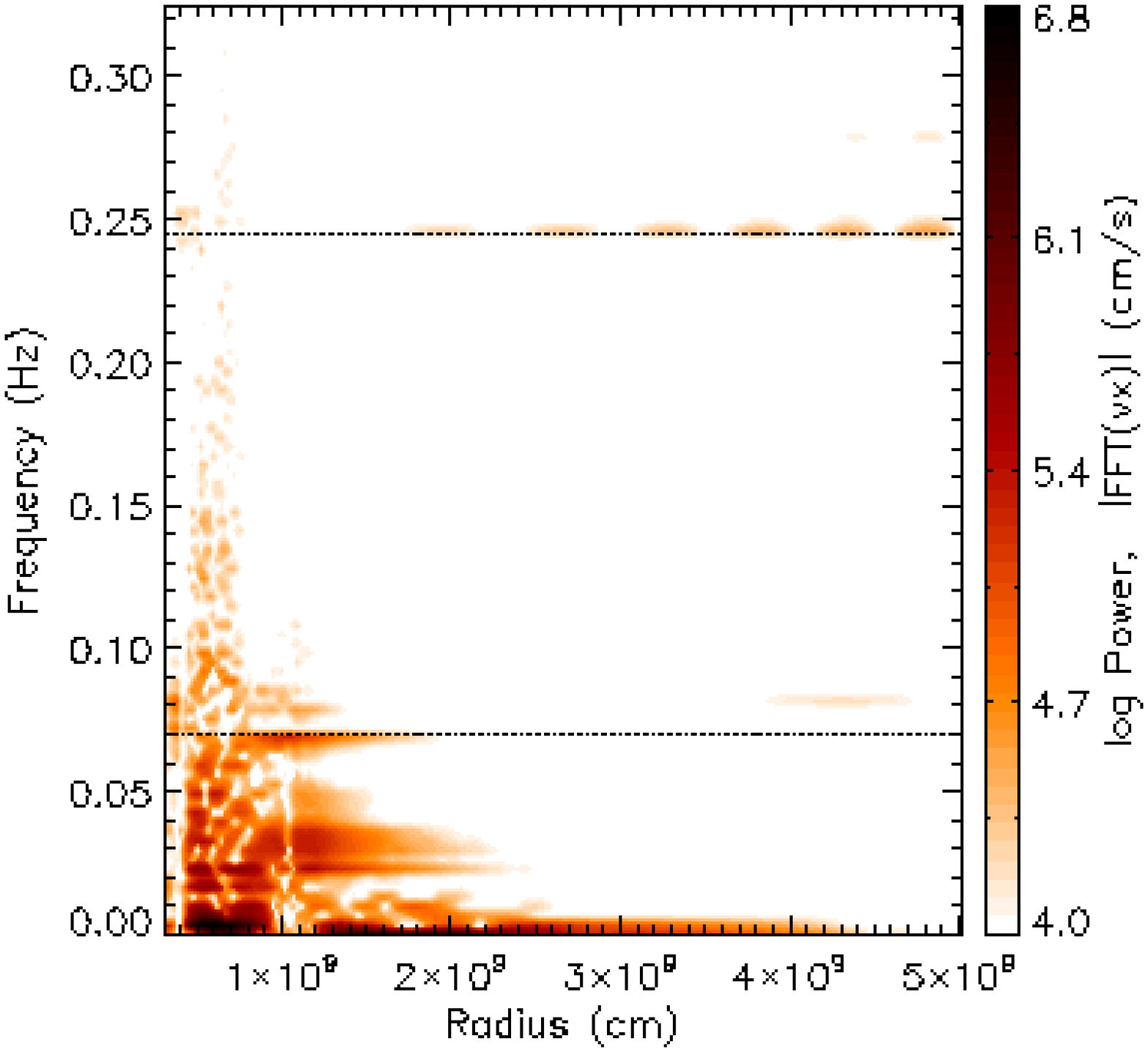}{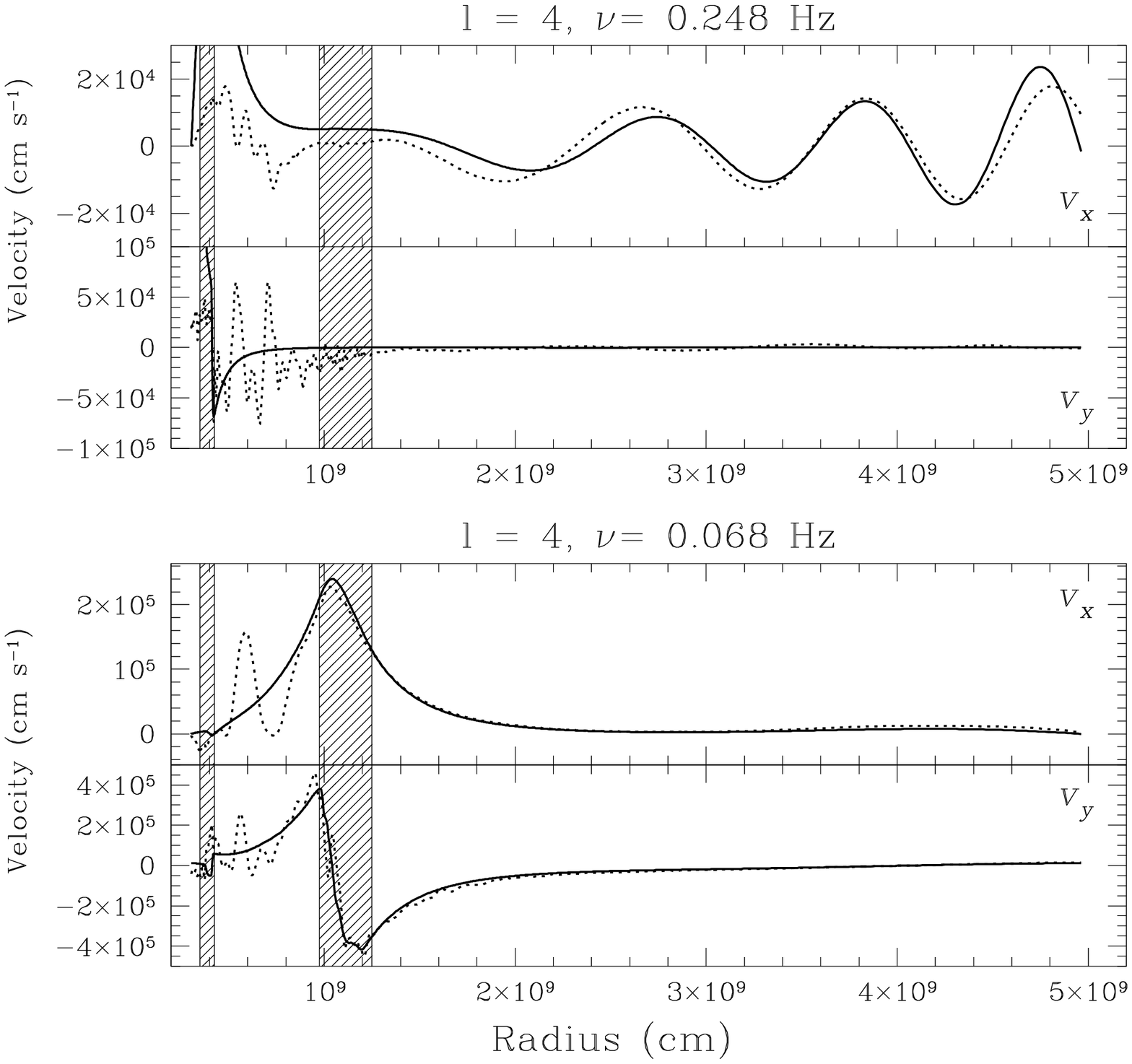}
  \caption{
    {\em (left)} The frequency power spectrum of the l=4 component of the radial 
    velocity ($v_x$) is shown as a function of radius.
    The dashed horizontal lines indicate the frequencies of two waves with significant
    power that are compared to linear theory in the adjacent panel.
    {\em (right)} Two l=4 wave forms are shown.
    The wave forms extracted from the simulation data (dotted line) and calculated with 
    linear theory (solid line) are shown for comparison.  For each mode the 
    horizontal ($v_y$) and radial velocity ($v_x$) components are presented in 
    units of cm s$^{-1}$.  The shaded  areas correspond to stably stratified regions.
    Both of these modes have some p- and g-mode character. The mode in the bottom panel
    is predominately of g-mode character, while the mode in the top panel is more
    clearly of mixed type.
    \label{f4}}
\end{figure*}

\par {\em Resonant Modes.}
The underlying stellar structure determines the set of discrete resonant modes 
that can be excited.  
The narrow stable layers which bound the convective shells in our simulation, including 
the (truncated) core layer, are isolated enough from other wave propagation regions to act 
as resonating cavities.  
These modes are the deeper, interior counterparts to the modes 
observed in helio and asteroseismology studies of milder evolutionary stages.

\par Each mode can be uniquely identified by its horizontal wavenumber index, $l$, 
and its oscillation frequency, $\omega$.  
We identify excited modes in our 
simulation by isolating spatial and time components of the motion
through Fourier transforms.  
In Figure \ref{f4} we present a power spectrum at each radius in the simulation
for motions with $l=4$, the largest horizontal scale that can fit into 
the 90$^\circ$ wedge simulated.  
A direct comparison between modes identified in the simulation and those calculated
from the linearized (non-radial) wave equation of stellar oscillations
\citep{unno1989} is presented in the right hand panel for two modes with 
significant power. 
Although the simulation data has additional features, including ``noise'' in the 
convection zones, the mode shapes in both velocity components are strikingly
similar between simulation and the wave equation; the identification is unambiguous.
Gravity waves evanesce (exponentially attenuate) 
beyond the boundaries of the stable layers but still contain 
significant power in the convection zones. Acoustic waves are free to 
propagate in the acoustic cavity which overlaps the carbon burning convection zone.

\par During the late evolutionary epoch simulated here, the g-mode and p-mode propagation 
zones are not widely separated in radius, allowing wave modes of mixed character to 
couple \citep{unno1989}.  The modes in the acoustic cavity are trapped by the boundary 
conditions of the calculation but would otherwise propagate into the stellar envelope 
where they would deposit their energy through radiative damping, providing an additional 
channel for energy transport out of the burning region.

\par The good agreement of the numerical modes with the analytic
modes indicates that our numerical procedures give an excellent
representation of the hydrodynamics of waves, even of very low mach number.
We note that anelastic codes will not reproduce the p-mode and
mixed mode waves properly, as we have ascertained by direct
integration of the anelastic wave equation.

\par{\em Kinetic Energies and Wave Induced Mixing.}
During the simulations, convective motions excite waves and build up
significant kinetic energy in nonconvective regions. The integrated kinetic
energies are $2.1 \times 10^{45}\rm ergs$ in the inner nonconvective
region, $5.2 \times 10^{47}\rm ergs$ in the oxygen burning convective
shell, $9.7  \times 10^{45}\rm ergs$ in the intermediate nonconvective
region, and  $6.9 \times 10^{46}\rm ergs$ in the carbon burning shell.
The kinetic energy is small in the outer stable region, but
is still increasing by the end of the simulation.

In our simulations, the importance of the excited g-modes in the stable layers lies
primarily in the role they play in mediating mixing at the convective boundaries and
in the stably stratified layers.  We identify a cycle in which kinetic energy builds up
in the stable layer until the underlying wave modes reach non-linear amplitudes, breakdown
and drive mixing.  This process is analogous to the physical picture which underlies
semiconvective mixing \citep{stevenson1979,langer1983,spruit1992} but is driven on a 
hydrodynamic rather than a thermal timescale.  The growth time for the fastest growing modes 
in our simulation is $\sim$200 seconds and leads to an average migration speed of 
outer oxygen shell boundary of $\sim$4$\times$10$^4$ cm s$^{-1}$, 
entraining mass into the convection zone at a rate of $\sim$10$^{-4}$ M$_{\odot}$ s$^{-1}$, 
significantly affecting the evolution.
Identifying the spectrum of excited modes in numerical simulations, including amplitudes 
and waveforms, provides guidance for developing and testing a quantitative model of this 
mixing mechanism.  An important parameter controlling the boundary entrainment rate is the
gradient Richardson number such that steeper density (and composition gradients) will lead 
to lower mixing rates so that sharp gradients are expected to form and persist 
\citep{peltier2003,alexakis2004}.  In addition to the mixing associated with wave-breaking,
enhanced compositional diffusion can be driven by the presence 
of the oscillatory flow setup by g-modes \citep{press1981b,knobloch1992}.  
It has been demonstrated that the structure of presupernova iron cores is very sensitive 
to how mixing is handled at convective boundaries with significant implications for both 
the explosion mechanism and nucleosynthetic yields \citep[e.g.][]{woosley1988}.
If the amplitudes of the wave motions identified in our simulations remain robust to 
the numerical limitations (e.g., resolution, domain size) then neglecting the 
mixing processes associated with these waves constitutes a large source of error
in progenitor models.

\section{Discussion}

\par {\em Differences in 2D and 3D.}
While these 2D simulations allow us to see the interaction of carbon
and oxygen shells, and show wave generation at convective boundaries,
they impose an unphysical symmetry on the problem. Our 3D simulations
have only been carried to $300$ seconds of stellar time, but show
that the wave generation is a robust result.
The flow in the convection zones, however, are qualitatively different between 
the 2D and 3D models: the 2D flow is dominated by vortices which span the convection zone,
while the 3D flow is characterized by smaller scale plumes. Quantitatively, the 
amplitude of the 2D convective motions are larger than the 3D by a factor of 8,
while the 3D motions are larger than mixing length values by a factor of 1.5. 
{\em Two-dimensional simulations of gravitational collapse will be
  misleading at least to the extent that convective motions are important.}

\par{\em Presupernova Models.}
Perhaps the most important impact that internal waves have on stellar structure in the
late stages of massive star evolution is the degree to which they drive
compositional mixing.  In our simulations
internal wave modes grow to non-linear amplitudes and
mix material at convective boundaries on a hydrodynamical timescale. 
Given the strong dependence of presupernova structure on the rate at which mixing
occurs at convective boundaries we see the incorporation of internal wave
physics into stellar evolution codes as a neccesary refinement.

\par{\em Symmetry Breaking.}
Spherical symmetry in presupernova models is broken by (1) the density
perturbations induced by turbulence within the convection zone,
(2) the wave interactions between burning shells, and (3) rotationally
induced distortions.
The perturbations by waves which are trapped between the oxygen 
and carbon burning shells are correlated on large angular scales, as
is rotation, while the turbulent perturbations have both a smaller scale and
amplitude. Our restricted simulation domain filters out wave modes with $l<4$,
so it is likely that even larger scale perturbations exist in real stars.
Symmetry breaking will seed instabilities in an outward propagating 
supernova shock \citep{kuranz2005}, and in the collapsing core. The converging
case has been intensely studied for inertial confinement fusion
\citep{lindl98}.  
The diverging case has implications for the problem of $\rm^{56}Ni$
and$ \rm^{56}Co$ decay in SN1987A \citep{herant1991,kifonidis2003}.

The conclusion that internal wave modes do not
grow to large amplitudes during core collapse through nuclear driven 
overstability \citep{murphy2004}, is based upon an analysis that
ignores (1) the dynamics of convective motion and (2) the shell-shell
interactions, both of which are expected to become more violent as collapse is
approached. Asymmetries in core collapse have implications for pulsar birth kicks,
explosion mechanisms, and for gravitational wave generation.
\citet{char05} have shown that internal gravity waves can transport
angular momentum at a rate sufficient to be important in the evolution
of solar mass stars; we suggest that they are important for
evolution of more massive stars to core collapse, and for plausible
prediction of the angular momentum distribution in that collapse.

\acknowledgements

The authors would like to thank Jeremiah Murphy, Christian Ott, Patrick Young,
Adam Burrows, Ed Olszewski, Luc Dessart and Philip Pinto for useful discussions
and comments.  We would also like to acknowledge the referee, Raph Hix,
for comments that have greatly improved the manuscript.
This work was supported in part by the University of Arizona and by a 
subcontract from the University of Chicago ASCI Flash Center.

\clearpage

\clearpage

\clearpage

\clearpage

\clearpage

\end{document}